\documentclass[twocolumn,showpacs,preprintnumbers,aps,prb]{revtex4}
\usepackage{graphicx}

\begin{document}

%
%

\title{Thickness dependence of linear and quadratic magneto-optical
Kerr effects in ultrathin Fe(001) films}

\author{M. Buchmeier,\cite{mbuch} R. Schreiber, D. E. B\"urgler, and
C. M. Schneider}
\affiliation{Institut f\"ur Festk\"orperforschung (IFF-9) and
JARA-FIT, Forschungszentrum J\"ulich GmbH, D-52425 J\"ulich, Germany}

\begin{abstract}
Magneto-optical Kerr effect (MOKE) magnetometry is one of the most
widely employed techniques for the characterization of ferromagnetic
thin film samples.  Some information, such as the magnitude of coercive fields or
anisotropy strengths can be readily obtained without any knowledge of the
optical and magneto-optical (MO) properties of the material. On the
other hand, a quantitative analysis, which requires a precise
knowledge of the material's index of refraction $n$ and the MO
coupling constants $K$ and $G$ is often desirable, for instance,
for the comparison of samples, which are different with respect to
ferromagnetic layer thicknesses, substrates, or capping layers.  While
the values of the parameters $n$ and the linear MO coupling parameter
$K$ reported by different authors usually vary considerably, the
relevant {\em quadratic} MO coupling parameters $G$ even for Fe are
completely unknown.  Here, we report on measurements of the thickness
dependence (0--60\,nm) of the linear and quadratic magnetooptical effects in epitaxial
bcc-Fe(001) wedge-type samples performed at a commonly used laser
wavelength of 670\,nm.  By fitting the thickness dependence we are
able to extract a complete set of parameters $n$, $K$,
$(G_{11}-G_{12})$, and $G_{44}$ for the quantitative description of
the MOKE response of bcc-Fe(001).  We find the parameters $n$, $K$, and
$G$ to significantly differ for films thinner than about 10\,nm as compared to
thicker films, which is indicative of a thickness dependence of the
electronic properties or of surface contributions to the MOKE. The
magnitude of the quadratic magnetooptical effect is found to be about one third of the
record values reported recently for Co$_2$FeSi.
\end{abstract}

%
\pacs{78.20.Ls 78.20.Ci 78.20.Bh 75.70.Ak 75.30.Gw 75.60.-d}

%

\date{\today}
\maketitle

%
%
\section{Introduction}
\label{intro}
%
%
The magneto-optical Kerr effect (MOKE) of ferromagnetic (FM) thin
films has been a field of intensive studies over the last three
decades. This interest in MOKE was motivated by three aspects: (i) its 
importance as a experimental magnetometric tool, (ii) as a means to 
measure the band structure of FM materials, and (iii) its application 
in magneto-optical (MO) storage media.
%
%
MOKE is probably the tool most widely employed for the magnetometric
characterization of thin-film samples employed for spintronics.  Among
its most common applications are the quantitative determination of the coercivity,
magnetic anisotropy, and interlayer exchange coupling from the
analysis of hysteresis loops recorded with the MOKE signal. Other prominent
applications are the investigation of spin dynamics in the time-domain
and magnetic domain imaging.  The main advantages of the MOKE over
other techniques are its compatibility with high magnetic fields, surface sensitivity 
with a typical information depth of some 10\,nm, a time resolution down to 
the sub-picosecond regime, a reasonable spatial resolution of the order of 
about 0.5\,$\mu$m, and robust and inexpensive experimental setups.

Many applications neglect the absolute magnitude of the Kerr effect, which
is given by the magnitude and phase of the complex Kerr angle.
Instead they describe the dependence of the normalized Kerr angle on
the magnetization angle by means of adjustable response coefficients.
This type of description has the advantage that it does not require
any knowledge of the materials's optics, yet it is sufficient to
extract a lot of information, such as the magnitude of the magnetocrystalline
anisotropy of FM single-layer systems,\cite{Shishen,JMMM_297_118} or
the antiferromagnetic interlayer exchange coupling of FM double-layer
systems.\cite{JMMM_240_235,PhD} On the other hand, the absolute magnitude of 
the complex Kerr angle provides valuable information, which can support the experimental
data, and can be employed to quantitatively compare samples with, for
instance, different FM layer thicknesses, substrates, or capping
layers and to determine the thickness of the FM layers, the MOKE
information depth, {\em etc.} However, the full quantitative MOKE information
is generally not linked by simple analytic formulae to the material
properties, which are the indices of refraction $n$ and the linear and
quadratic MO coupling parameters $K$ and $G$ of all involved
layers.  Even in ultrathin films of only some nanometers thickness, a linear
dependence of the size of the MOKE response on the FM layer thicknesses, known
as additivity law -- which has been claimed by Qiu {\em et al.} in
Ref.\,\onlinecite{PRB_45_7211} -- is generally not valid.\cite{JAP_98_033516}

Therefore, a general numerical treatment of the MOKE by solving
Maxwell's equations and the standard boundary conditions is
indispensable for the quantitative interpretation of the Kerr angle.  A
prerequisite for this calculation is the precise knowledge of the
optical and magnetooptical material parameters.  Although spectroscopically determined values for
$n$ and $K$ are available for many materials, the overall agreement of
the data from different sources is often, as for instance in the case
of bcc-Fe,\cite{Oppeneer} rather poor.  The strong variation of the
tabulated optical constants is frequently attributed to surface
contamination or oxidation of the {\em ex-situ} measured samples, but
thickness or quality variations of the films are also plausible.  For instance,
(i) in Ref.\,\onlinecite{PRL_80_5200} the Kerr angle of Fe has been
found to oscillate as a function of the thickness of a capping Au
layer due to quantum well states, (ii) spectroscopic MOKE data of thin
Fe films show features, which cannot be explained by the bulk
electronic band structure of Fe,\cite{JAP_91_8246} and (iii) there is
strong evidence for interfacial contributions in the
MOKE.\cite{PRB_64_155405}

Moreover, the literature values are almost exclusively limited to the
first order linear MO coupling parameter $K$. For
many FM materials, {\em e.g.} from elementary Fe
\cite{JMMM_172_199,PRB_55_8990,Shishen,JAP_91_7293} to the more
complex Heusler alloy Co$_2$FeSi,\cite{JPD_40_1563}, however, the second order
quadratic coupling constants $G$ have a comparable impact on the
Kerr angle.  To our knowledge the only published value of a second
order MO coupling constant is the imaginary part of $G_{44}/K$ of Ni
of about $Im(G_{44}/K)=-0.02$ at the wavelength $\lambda=514$\,nm,
which has been experimentally determined from Brillouin light
scattering data by Giovannini {\em et al.}\cite{PRB_63_104405} The
second order magnetooptical parameters, which are $G_{11}$, $G_{12}$, and $G_{44}$
for systems with cubic symmetry, give rise to MO effects quadratic and
even in the magnetization $M$.  These effects are known as quadratic
MOKE (QMOKE) or Voigt effect in reflection, and have recently received 
a lot of attention.\cite{JMMM_172_199,PRB_55_8990,Shishen,JAP_91_7293,JPD_40_1563}
Effects quadratic in $M$ also turn out to be important in magnetization-dependent second
harmonic generation (MSHG),\cite{PRB_75_64401} x-ray magnetic linear
dichroism and the closely related x-ray Voigt effect.\cite{XMLD} All
the more it is surprising, that to our knowledge no one has yet
determined a full set of corresponding material parameters $G$ to
describe the QMOKE.

%
%
Apart from their practical significance for MOKE magnetometry, the MO
coupling parameters are also important in fundamental research.  From
a microscopic point of view the MO coupling is due to the interplay
of the exchange exchange interaction leading to a splitting of the 
bands and the spin-orbit (SO)
coupling. It is therefore closely related to the magnetocrystalline
anisotropy energy, which also arises due to SO coupling. The MO
coupling parameters may be seen as an important probe for the fundamental
electronic interactions in FM materials, {\em e.g.} spectroscopic MOKE
is a widely used standard tool to evaluate the band structure in FM
metals.\cite{Oppeneer} While the linear MO effect is due to first
order SO coupling, the quadratic MO coupling is thought to be caused
by second order SO coupling terms.  The SO coupling is known to be
altered at the interfaces and in ultrathin films, which gives rise to
the well-known thickness dependence of the magnetocrystalline
anisotropy energy, {\em e.g.} in Fe,\cite{MAE} and interfacial
MOKE contributions, which have been observed by various
authors.\cite{PRB_64_155405}

%
%
In this contribution we report on a magnetometric study of the 
magnetooptical response of bcc-Fe(001)
wedge-type samples with thicknesses ranging from 0 to 60\,nm.  We have
determined both components of the complex Kerr angle, the Kerr
rotation and the Kerr ellipticity.  Effects linear and quadratic in
$M$, LMOKE and QMOKE respectively, are separated by fitting the
hysteresis loops to a single domain model.  The QMOKE, which is known
to be anisotropic, {\em i.e.} it depends on the sample orientation
with respect to the plane of incidence, has been determined for both
Fe(001)[110] and Fe(001)[100] directions parallel to the plane of
incidence.  By fitting the thickness dependence of LMOKE and QMOKE we
are, for the first time, able to extract a full set of Fe material
parameters $n$, $K$, $(G_{11} - G_{12})$, and $G_{44}$ at a light
wavelength of 670\,nm.  We find a sizeable thickness dependence,
which, however, seems to be mainly of non-magnetic origin.  The main
effect is an increased index of refraction for Fe film thicknesses
below about 10\,nm as compared to thicker films.  A large maximum
absolute value of the quadratic Kerr effect (QMOKE) of 0.37\,mrad is
found at about 22\,nm Fe thickness.  This value is of a comparable 
order of magnitude as the recently reported record QMOKE values for 
Co$_2$FeSi.\cite{JPD_40_1563}

The paper is organized as follows: In Sec.\, \ref{prep} we describe
briefly the sample preparation.  Details of the experimental MOKE
setup and the data recording are given in Sec.\,\ref{setup}.
Section\,\ref{model} deals with the modeling of the MOKE and the
hysteresis loops.  The results are presented and discussed in
Sec.\,\ref{results}.  Finally, we summarize our results in
Sec.\,\ref{conclusions}.

\section{Experimental details}
\label{exp}
\subsection{Sample preparation and experimental MOKE setup}
\label{prep}
Epitaxial Fe(wedge)/Ag(1\,nm)/Au(2\,nm) have been prepared by
molecular beam epitaxy on top of a GaAs/Ag(001) buffer system.  The Au
capping layer has has been chosen thick enough to prevent oxidation
and thin enough to be able to determine large Kerr angles.  The Ag
interface layer has been introduced in order to prevent a possible alloying
of Fe and Au.  The preparation is described in detail
elsewhere.\cite{BUER96-1,BUER97-1} All thicknesses have been precisely
determined using a calibrated quartz crystal monitor.  The Fe
thickness has been varied continuously between 0 and 8\,nm for sample
A and stepwise in sample B with discrete Fe thicknesses of 5, 8, 12,
18, 24, 32, 44, and 60\,nm.

\begin{figure}[b]
\centering\leavevmode
\includegraphics[width=0.9\linewidth,clip]{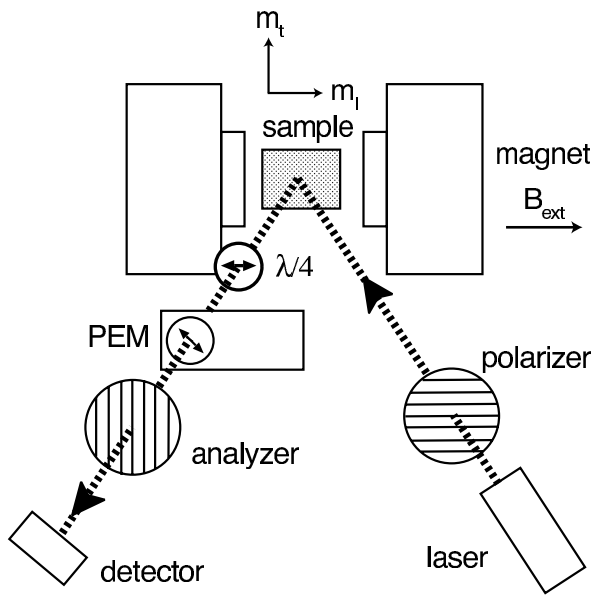}
\caption{Experimental MOKE setup.  The quarter wave plate is
introduced to measure the Kerr rotation $\theta$ instead of the Kerr
ellipticity $\epsilon$.  For clarity, the diaphragm is not shown.}
\label{f-setup}
\end{figure}

\subsection{MOKE setup}
\label{setup}
The MOKE measurements were performed using light from a Toshiba TOLD9231M
multi-mode laser diode with a wavelength of 670\,nm and a spectral
half-width of less than 2\,nm.  An in-plane
magnetic field with a maximum strength of 0.7\,T is generated by a
Broker electromagnet with FM yoke and measured with a Hall probe.  The
sample is mounted on a manually rotatable holder with an angle scale,
which allows a determination of the angle with respect to the plane of
incidence with a precision of about $\pm2^{\circ}$.

The optical setup consists of a light beam with its plane of incidence
parallel to the external field direction illuminating the sample under
an angle of incidence of 15$^{\circ}$ with respect to the sample
normal (see Fig.\,\ref{f-setup}).  The incident light is polarized in
$\hat p$ direction (electric field component in the plane of
incidence) employing a rotatable Glan-Taylor type polarizer.  The
phase of the reflected light is modulated at a frequency $f=50$~kHz
with a HINDS PEM-90 photo elastic modulator (PEM) at diagonal
modulation axis (rotated by 45$^{\circ}$ with respect to the $\hat
p$-direction) and at a retardation of 90$^\circ$.  A quarter wave
plate with its retardation axis parallel to the $\hat p$ direction
can be placed between sample and PEM. Consecutively the light passes
through an analyzer of the same type as the polarizer, but oriented in
$\hat s$ direction.  The light intensity is then converted into an
electrical voltage by a homemade diode detector.  In order to avoid
possible multiple reflections between the sample and the optical
elements in the reflected beam-path, which might impair the data by
parasite signals, a diaphragm with a diameter of about 2\,mm is placed
right after the sample.  Analyzer, PEM, and quarter wave plate are
slightly tilted with respect to the optical axis so that the light
reflected back to the sample is blocked by the diaphragm.

With this setup the small $f$ (50\,kHz) component determined with a
Lock-In amplifier is to first order proportional to the $\hat
p$-ellipticity $\epsilon$ times the reflected intensity, while the
much larger $2f$ (100\,kHz) and DC components, recorded simultaneously
with a multimeter are to first order proportional to the reflected
intensity alone.\cite{diplom} By introducing a quarter wave plate
between sample and PEM we are able to measure the $\hat p$-rotation
$\theta$ instead of the ellipticity.  The two Kerr angle components
($\theta$ and $\epsilon$) are calculated by dividing the measured $f$
component by the $2f$ component.

%
As the amplification factor of the detector is frequency-dependent,
the measured Kerr angle has to be calibrated.  For this purpose, we have used a
thick Au sample and turned the analyzer out of the $\hat p$ direction
by an angle $\Psi$, while recording the ellipticity and rotation
signal.  The data are then adjusted such that the measured dependence
of the resulting rotation $\theta$ and ellipticity $\epsilon$ as a
function of the polarizer angle $\Psi$ matches the theoretical
relationship
\begin{equation}
\theta(\Psi) +i
\epsilon(\Psi) = \frac {r_{ss}} {r_{pp}} \sin(\Psi),
\label{e-cali}
\end{equation}
where $r_{ss}$ and $r_{pp}$ are the diagonal Fresnel reflection
coefficients.  A value $r_{ss}/r_{pp} = 1.0003 - 0.0379 i$ has been
calculated using the Fresnel formulas and taking the literature
value \cite{CRC_HANDBOOK} $n = 0.100+3.653 i$ for the index of
refraction of Au at our laser wavelength.  We have performed the
calibration procedure for both the real and imaginary parts of
Eq.\,(\ref{e-cali}), {\em i.e.} with and without quarter wave plate,
and find an excellent agreement for the calibration factors (less than
2\% difference).  This corroborates that the calibration works
properly and possible detriments of the measurement by multiple
reflections are indeed well under control.

\section{Numerical modelling of the MOKE and hysteresis loops}
\label{model}
The optical and MO material properties can be described by the
permittivity tensor $\epsilon _{ij}$, which can be expanded in a power
series of the Cartesian direction cosines $m_i$ of the magnetization
$\vec{M}$:
\begin{equation}
\epsilon_{ij}(\vec{M}) = \epsilon_{ij} ^{(M=0)}
          + K_{ijk} m_k + G_{ijks} m_k m_s      +\cdots .
\end{equation}
The number of independent linear and quadratic MO coupling constants,
$K_{ijk}$ and $G_{ijks}$ respectively, is reduced by the symmetry of
the crystal and the Onsager principle $\epsilon_{ij}(\vec{M}) =
\epsilon_{ij}(- \vec{M})$.  For cubic symmetry the permittivity tensor
is completely defined by five quantities: The nonmagnetic part of
the permittivity $\epsilon ^{(M=0)} = n ^2$, which is given by the
square of the index of refraction $n$, the linear MO coupling constant
$K = K_{ijk}$, and three independent quadratic MO coupling parameters
$G_{iiii} = G_{11}$, $G_{iijj} = G_{12}$ and $G_{ijij} = G_{44}$.
Instead of the linear MO coupling $K$, the Voigt parameter $Q = i K m
/\epsilon ^{(M=0)}$ is frequently used.  The complex Kerr angle, which
is a measure of the magnitude of the MOKE signal can be calculated using the
standard $4\times4$ matrix formalism as explained in
Refs.\,\onlinecite{Yeh,CJP_41_663,PRB_43_6423}.  Our open-source computer program developed for the calculation of the MOKE can be
downloaded from Ref.\ \onlinecite{code}.

The small complex Kerr angle is generally defined as
the off-diagonal divided by the diagonal Fresnel reflection
coefficients, for instance for incident p-polarized light:
$\Phi _p = r_{ps}/r_{{pp}} = \theta +i\epsilon$.
%
%
We would like to point out that this definition is insufficient as it
does by no means define the {\em sign} of the Kerr angle, which
depends on the choice of coordinate system, relative orientation of
the $\hat p$ and $\hat s$ directions, the in-plane wave-vector of the
light, and the sign of the exponent of the wave function.  The sign
convention in magneto-optics is indeed a long-standing problem, {\em
i.e.} different authors report different signs for the complex Kerr
and Faraday rotation angles,\cite{Oppeneer}. Even worse hardly any
article gives a clear definition of the employed sign conventions.
Therefore, we will here briefly derive an unambiguous definition of the
sign, which must include the rotational sense of the Kerr angle, the
geometry of the experimental setup, and the orientation of the
longitudinal and transversal components $m _l$ and $m _t$ of the
magnetization.  We define the complex rotation to be positive when the
rotational vector is pointing in the propagation direction of the
reflected light, {\em i.e.} the polarization vector ($\hat p$
direction) is rotated in clockwise direction when looking in the
direction of the reflected beam.
%
%
The orientation of the coordinate system is defined as depicted in
Fig.\,\ref{f-setup} with positive $m_l$, which is also the
direction of a positive external field pointing to the right, $m
_t$ pointing up, and $k _\parallel$, which is the direction of the
in-plane light wavevector, pointing to the left when looking onto the
the sample.  With this convention for a 60\,nm Fe film the Kerr
rotation due to a positive $m_l$ results in a negative
Kerr rotation $\theta$ and a positive Kerr ellipticity $\epsilon$.
The negative Kerr rotation is equivalent in sign to a
counterclockwise turn (looking in direction of the incident beam) of
the polarizer and the positive ellipticity $\epsilon$ to a
clockwise turn of the polarizer out of the $\hat p$ direction,
respectively.

It is convenient to expand the Kerr angle $\Phi$ as a function of the
directional cosines of the magnetization vector $\vec
M$,\cite{JAP_91_7293} {\em e.g.} for in-plane magnetization:
\begin{equation}
\label{phenMOKE}
\Phi
= \sum _{layers~i} \Bigr[ l _i m _{l, i} + q _{1, i} m _{l, i} m _{t, i}
+ q _{2, i} m _{t, i} ^2 +t _i m _{t, i} +O(m^3) \Bigl],
\end{equation}
where $l _i$ are the longitudinal (LMOKE), $q _{1, i}$ and $q _{2, i}$
the quadratic, and $t _i$ the usually much smaller transversal (TMOKE)
response coefficients.  Similar relations hold for the more general
case including out-of-plane magnetization, the Faraday effect, and
even the calculation of Brillouin light scattering intensities,
\cite{PRB_75_184436} which are closely related to the MOKE problem.

%
%
Note that some authors use a different form for the second quadratic
term, namely $q _2 (m _l ^2 - m _t ^2)$ 
, which is equivalent 
to our term, if the magnetization is in a single domain state
except for a factor of $-2$ and a constant offset of $1$.  
In the case of a multi-domain state this kind of
description does not generally hold as additional significant MO
effects due to the magnetization gradient \cite{PSSA_118_271} can be
present.  The longitudinal coefficients $l_{i}$ stem from the linear
MO coupling parameter $K$ alone and are known to be isotropic, {\em
i.e.} independent on the sample orientation, as long as the FM layers have
cubic symmetry.
By contrast, the quadratic coefficients are due to a combined effect
of the linear and quadratic MO couplings and are anisotropic, {\em i.e.}
they depend on the relative orientation of the sample with respect to
the plane of incidence.  For cubic systems the resulting $q$
coefficients have been found to have the form: \cite{JAP_91_7293}
\begin{eqnarray}
\label{QMOKEanisotropy1}
q _1 = q _{001} + (q _{011} - q _{001}) \sin^2(2 \gamma)  \\
\label{QMOKEanisotropy2}
q _2 = \frac 1 2 (q _{011} - q _{001}) \sin(4 \gamma),
\end{eqnarray}
where $\gamma$ is the angle between the in-plane component of the
light wavevector and an in-plane Fe[001]
direction, and $q _{001}$ and $q _{011}$ are QMOKE constants for the
plane of incidence parallel to the [001] and [011] directions,
respectively.  The anisotropy of the QMOKE stems from the symmetry of
the effective SO coupling tensor $G_{ijks}$, which is closely related
to the symmetry of the crystal.  For Fe(001) with in-plane
magnetization $(G_{11} - G_{12})$ and $2G_{44}$ are the relevant SO
coupling parameters for the [011] ($\gamma=45^\circ$) and [001]
direction ($\gamma=0^\circ$), respectively.  The parameter $\Delta G =
(G_{11}-G_{12})-2G_{44}$ is a measure for the anisotropy
strength.\cite{JAP_91_7293}

The $q _1$ coefficients have an isotropic and an anisotropic
contribution.  They depend on the sum of $-K^2/n^2$ 
and $G_{11}-G_{12} - \Delta G \cos ^2 (2\gamma)$.
\cite{JAP_91_7293} Therefore, errors in the determination of $n$ and
$K$ affect the accuracy of the $G$ values.  A wrong sign of the $q _1$
value will not simply lead to wrong signs, but to wrong values of the
second order MO coupling constants.  On the other hand, the $q _2$
coefficients stem purely from the $G$ parameters, namely from $\Delta
G /2 \sin (4\gamma)$,\cite{JAP_91_7293} and vanish if the plane of
incidence is parallel to the symmetry directions [001] and [011].

The response coefficients can be determined experimentally -- at least
for single layer systems -- with a suitable setup, for instance, by rotating
the field.  \cite{JAP_91_7293,ROTMOKE} They can also be
calculated numerically from the optical and MO material parameters $n$, $K$, and
$G$.  As our MOKE setup does not allow for a field rotation, we choose
here the alternative approach of analyzing remagnetization loops
recorded at different sample orientations.

\begin{figure}[hbt]
\centering\leavevmode
\includegraphics[width=0.9\linewidth,clip]{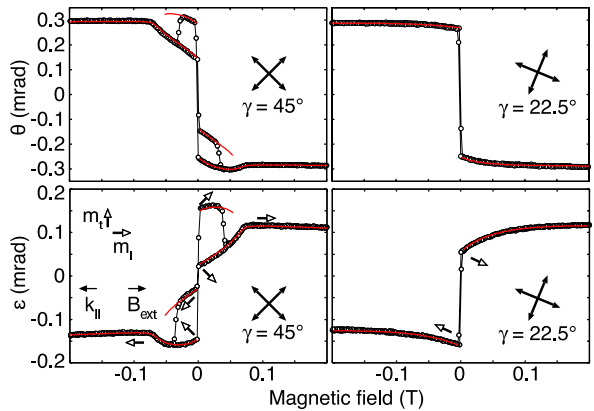}
\caption{Field dependence of measured (open circles) and
calculated (red lines) Kerr ellipticity (bottom graphs) and Kerr
rotation (top graphs) for a 60\,nm Fe film.  The left curves
correspond to a sample orientation with hard axis parallel to external
field ($\gamma \approx 45^{\circ}$), while for the right curves the
field is parallel to an intermediate direction ($\gamma \approx
22.5^{\circ}$).  The field is swept form positive to negative values.
The magnetization alignments are show with short arrows, the easy axis
directions of the magnetocrystalline anisotropy with crossed long
arrows.}
\label{fig_loops}
\end{figure}

Typical MOKE loops for a 60\, nm Fe film are shown in
Fig.\,\ref{fig_loops}.  The experimental Kerr rotation (top graphs)
and ellipticity (bottom graphs) are plotted with connected open
circles.
%
%
We have determined the MOKE response coefficients $l$, $q_1$, and
$q_2$ by fitting experimental remagnetization loops to a single domain
model taking into account the sample orientation $\gamma$, the cubic
anisotropy parameter $K_c/M_s$, and describing the Kerr angle via
Eqs.\ (\ref{phenMOKE}--\ref{QMOKEanisotropy2}).  While the left loops
recorded at an angle $\gamma \approx 45^\circ$, {\em i.e.} with field
parallel to a hard [011] direction, depend on $l$ and $q _{011}$, the
right loops are recorded at $\gamma \approx 22.5^\circ$ and,
therefore, depend on $l$ and both $q _{011}$ and $q _{001}$.  Thus, a
simultaneous fitting of the loops for both orientations yields a full set
of MOKE response coefficients $l$, $q_{001}$, and $q_{011}$.


\section{Results and discussion}
\label{results}
\subsection{Linear MOKE}
\begin{figure}[hbt]
\centering\leavevmode
\includegraphics[width=0.9\linewidth,clip]{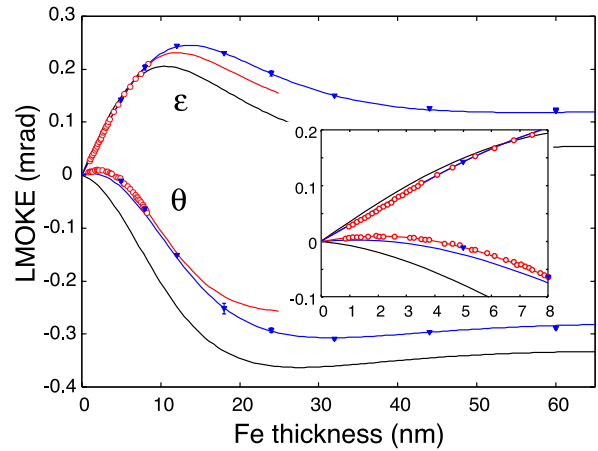}
\caption{Thickness dependence of the measured (red circles and blue
triangles for sample A and B, respectively) and calculated (lines)
linear ellipticity $\epsilon$ (upper curves) and rotation $\theta$
(lower curves).  Error bars are comparable to the symbol size.  Black
curves result from a calculation with material parameters from
literature.  A fit to the data of sample A with thickness range 0 -
8\,nm (red curves) yields a clearly different behavior than a fit to
the data of sample B with a larger thickness range 5 - 60\,nm (blue
curves).  Inset: Magnification of the low thickness region.}
\label{fig_lmoke}
\end{figure}

The thickness dependence of the experimental linear MOKE extracted
from the hard axis loops is marked in Fig.\ \ref{fig_lmoke} by red
circles and blue triangles for sample A and B, respectively.  The data
for thicknesses of 5\,nm and 8\,nm measured for both samples are in
excellent agreement indicating that the sample quality and thickness
calibration of samples A and B are very similar. The upper and lower
curves corresponds to the imaginary and real part of the Kerr angle
(ellipticity $\epsilon$ and rotation $\theta$), respectively.  In
agreement with the results of previous publications on similar
systems, \cite{PRB_45_7211} the Kerr ellipticity increases linearly with the film
thickness below 5\,nm, which is indicative for the law of additivity
of the MOKE effect size to be valid in this regime.  On the other hand, there is no such linear
behavior for the Kerr rotation, which changes sign at about 4\,nm,
meaning that the additivity does not hold for the Kerr rotation.  This
breakdown of the additivity law is due to the dominant imaginary
phase of the Kerr angle found for ultrathin Fe layers on noble metal
substrates.  In contrast, on semiconducting substrates, {\em e.g.}
GaAs, the Kerr angle of Fe is mainly real and the
additivity holds for the rotation, but not for the ellipticity.
\cite{JAP_98_033516} For thicker layers, the phase of the
electromagnetic wave inside the Fe layer changes due to the real part
of the perpendicular wavevector component $k_{\perp}$ and therefore,
gives rise to negative Kerr ellipticity contributions coming from Fe layers
buried deeper than about 12\,nm, where the slope changes sign.  On the
other hand, the imaginary part of the wavevector $k_{\perp}$ leads to
a decreasing intensity of the electromagnetic wave with increasing
depth inside the film, which determines the information depth of
about 40\,nm, where the slope begins to asymptotically flatten.

The thickness dependence calculated using literature values of the
indices of refraction from Refs.\ \onlinecite{CRC_HANDBOOK} and
\onlinecite{Yolken}, $n_{Ag} = 0.27 + 4.66 i$, $n_{Au} = 0.10 + 3.65
i$, $n_{Fe} = 3.57 + 4.02 i$, and the linear MO coupling from Ref.\
\onlinecite{Krinchik} $Q = 0.0437 +0.0040 i$ is plotted as black
lines.  As found earlier by Qiu {\em et al.}, \cite{PRB_45_7211} the
material constants from literature approximately reproduce the Kerr
ellipticity, which is insensitive to small phase changes of $Q$.
However, the literature data fail to describe the Kerr rotation.

We have fitted our experimental data employing the full $4\times4$
matrix formalism \cite{Yeh,CJP_41_663} using fixed indices of
refraction for Ag and Au from literature as specified above, and
treating the index of refraction and the MO coupling of Fe as free
parameters.  The red and blue curves in Fig.\ \ref{fig_lmoke}
correspond to the data of samples A and B, respectively.  The fit
results are listed in Table \ref{table_results}.  It turns out that a
satisfactory fit over the whole thickness range with only one
thickness-independent set of material parameters is impossible.  The
fit to the data of sample A with smaller thicknesses results in a
significantly about 10\% larger index of refraction with different
phase as compared to the thicker sample B. On the other hand, the MO
coupling parameter $Q$ mainly differs in phase by about 10$^{\circ}$,
while the absolute values are in rather good agreement within less than
3\% difference.  Thus, it seems that the difference between thin and
thick Fe layers is mainly of optic rather than of magneto-optic
origin.

Our value of the index of refraction of Fe determined from the data of
the thicker sample B, $n _{Fe} = 3.53 + 3.72 i$, compares reasonably
well with the value $n = 3.57 +4.02 i$ of Yolken and
Kruger.\cite{Yolken} While the real part is in excellent agreement,
our imaginary part is about 7\% smaller, which is probably within in
the range of the systematic experimental errors.
On the other hand, our value for the linear MO coupling $Q$ for sample
B is significantly by about 20\% smaller and has a phase difference
of about 17$^\circ$ compared to the data of Krinchik and
Artemev.\cite{Krinchik}

The curve fitted to the data from the thinner
sample departs from the experimental data at about 15\,nm, which
corresponds to approximately half the penetration depth of the light.
This circumstance might hint at an improper description of the optical
properties of the Ag substrate as a reason for the apparent thickness
dependence of the index of refraction of Fe.  The substrate mainly
influences the Kerr angle for Fe thicknesses below 15\,nm as the light
reflected from the substrate can reach the sample surface. In order
to test this conjecture, we have additionally fitted the data with the
indexes of refraction of the substrate and capping layers as free
parameters.  However, we could not substantially improve the overall
quantitative agreement of the fits.  Therefore, improper optical
parameters of the non-magnetic layers can be ruled out as a reason for
the encountered thickness dependence of the optical Fe properties.
Possible explanations for the thickness dependence are: (i) The
tensile strain of the Fe due to the small lattice mismatch of 0.7\%
between Fe and the Ag substrate leading to an anisotropic permittivity
tensor, (ii) altered electronic properties of the thin Fe layer due to
the proximity to the noble metal substrate and the capping layers,
which can have a sizable influence,
\cite{PRL_80_5200,PRB_64_155405} and (iii) interfacial MOKE
contributions, \cite{PRB_64_155405} which have been neglected in the
theoretical description.

\subsection{Quadratic MOKE}
\begin{figure}[hbt]
\centering\leavevmode
\includegraphics[width=0.9\linewidth,clip]{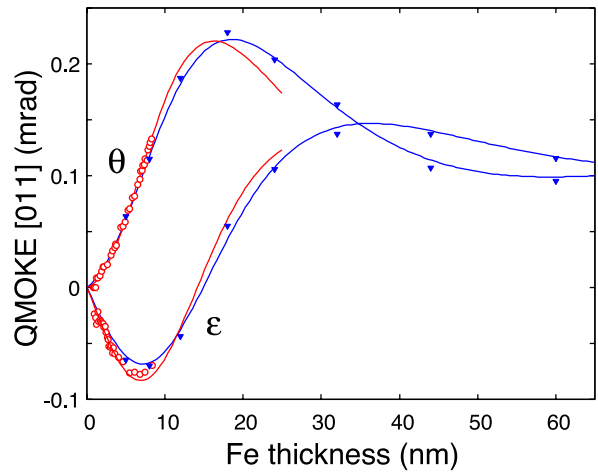}
\caption{Thickness dependence of the measured (red circles and blue
triangles for sample A and B, respectively) and calculated (lines)
QMOKE for the Fe[011] direction (hard axis) parallel to the plane of
incidence.}
\label{fig_qmoke_011}
\end{figure}
\begin{figure}[hbt]
\centering\leavevmode
\includegraphics[width=0.9\linewidth,clip]{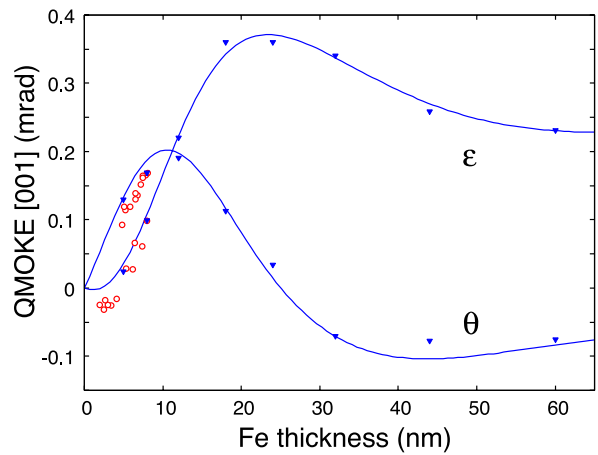}
\caption{Thickness dependence of the measured (red circles and blue
triangles for sample A and B, respectively) and calculated (line)
QMOKE for the Fe[001] direction (easy axis) parallel to the plane of
incidence.}
\label{fig_qmoke_001}
\end{figure}
The thickness dependence of the QMOKE coefficients $q_1$ for the hard axis
(Fe[011] direction) and the easy axis (Fe[001] direction) configurations
are plotted in Figs.\ \ref{fig_qmoke_011} and \ref{fig_qmoke_001},
respectively.  In contrast to LMOKE, in this case the real part (Kerr rotation
$\theta$) is the quantity, which depends linearly on thickness for
ultrathin Fe layers, while the slope of the imaginary part (Kerr
ellipticity $\epsilon$) changes sign at approximately 7 and 2\,nm for
the [011] and [001] configuration, respectively.  Thus, taking into
account the significant QMOKE contribution the additivity holds for
neither phase of the Kerr angle.

The extracted relevant second order MO coupling constants, $G_{11} -
G_{12}$ and $G_{44}$, for the hard and easy axis configurations are
listed in Table\ \ref{table_results}.  They have been determined by
fitting the data taking into account the $n$ and $Q$ parameters from
the fits to the LMOKE data (Fig.\ \ref{fig_lmoke}).  The determination
of the $q_{001}$ parameter depends on the both the values of $q_{011}$
and $l$ [see Eq.\ (\ref{phenMOKE})].  Therefore, the propagation of
errors results in a significantly poorer quality of the experimental
data.  As a consequence we could not reliably determine the $G_{44}$
parameter for sample A.

\begin{table*}
\caption{\label{table_results}Index of refraction ($n$), linear ($Q$)
and quadratic ($G_{11} - G_{12}$, $G_{44}$) MO coupling constants
derived from data of samples A and B.}
\begin{ruledtabular}
\begin{tabular}{c|c|c|c}
Parameter & Sample A (0 - 8\,nm)  & Sample B (5 - 60\,nm)  & Literature
  \cite{CRC_HANDBOOK,Yolken}\\
\hline
$n$ & $4.06 \pm 0.03 + (3.85 \pm 0.03)i$ & $3.53 \pm 0.03 +(3.72 \pm 
0.03)i$ & $3.57 + 4.02i$ \\
$Q$ & $0.0331 \pm 0.0002 - ( 0.0127 \pm 0.0002)i$ & $0.0356 \pm 
0.0004 -(0.0074 \pm 0.0003)i$ & $0.0437 + 0.004i$ \\
$G_{11} - G_{12}$ & $-0.0544 \pm 0.0005 -(0.0287 \pm 0.0005)i$ & 
$-0.0358 \pm 0.0002 -(0.0382 \pm 0.0002)i$ & \\
$G_{44}$ & & $-0.0117 \pm 0.0003 -(0.0349 \pm 0.0003)i$ & \\
\end{tabular}
\end{ruledtabular}
\end{table*}

We find maximum absolute values $|q_1|$ of the QMOKE at about 22\,nm
of 0.23\,mrad and 0.37\,mrad for the hard and easy axis
configurations, respectively.  These values should be compared to the
record QMOKE value of about 1.05\,mrad \cite{note} recently found in
Co$_2$FeSi alloys in Ref.\ \onlinecite{JPD_40_1558}.  This comparable
order of magnitude of the QMOKE of Fe and Co$_2$FeSi implies that the
maximum QMOKE value of Fe in the visible wavelength region might be
even larger than that of Co$_2$FeSi as both materials are expected to
have a distinct frequency dependence resulting from their electronic
band structures.

In Fig.\ 2 of Ref.\,\onlinecite{JAP_91_7293} Postava {\it et al.} give
the dependence of LMOKE and QMOKE contributions on sample orientation
for a 50\,nm thick bcc-Fe(001) sample capped with 1.5\,nm Pd and
measured at an incident angle of 3.25$^{\circ}$.  Based on our fitted
optical and MO material constants we have calculated the $Re(l)$,
$Re(q_{001})$, and $Re(q _{011})$ constants for the sample structure
and experimental configuration of Ref.\,\onlinecite{JAP_91_7293} and
assuming $n_{Pd} = 1.87 + 4.44 i$.  We find a value of $Re(l)= -0.066$\,mrad
which is comparable, but about 20\% smaller than the value of $Re(l)
\approx -0.09$\,mrad\cite{note} determined by Postava {\em et al.} On
the other hand, our QMOKE data differ more distinctly, although they
are of a similar order of magnitude.  While we find $Re(q_{011}) =
0.079$, which is about 30\% larger than the value $Re(q_{011}) \approx
0.06$ determined by by Postava {\em et al.}, our value for
$Re(q_{001}) = -0.073$ is about a factor of 2 smaller than
$Re(q_{001}) \approx -0.16$ determined by Postava {\em et al.} This
difference of the QMOKE is clearly larger than the experimental errors
caused by uncertainties in the layers thicknesses, the optical
properties of the capping layers, and the calibration of the MOKE
setup.  Although both the sample of Postava {\em et al.} and our
samples are epitaxial bcc-Fe(001), a possible explanation could be a
strong structural dependence of the QMOKE as described in Ref.\
\onlinecite{JPD_40_1563} for Co$_2$FeSi.
%


\section{Conclusions}
\label{conclusions}
The thickness dependence of the linear and quadratic MOKE of
wedge-type thin Fe(001) films magnetized in the sample plane has been
measured.  Good quantitative agreement of the experimental data with
calculations assuming bulk-type optic and MO material constants
indicates that the thickness dependence of the MOKE is mainly due to
bulk-type magnetooptical coupling.  On the other hand, we found a sizable
departure from theory for thicknesses below about 10\,nm.  This can be
explained by thickness dependent optic and MO material parameters,
which are possibly due to MO surface effects or thickness dependent
features of the bandstructure, {\em e.g.} quantum-well states.  By fitting the experimental data to results of a numerical model, we
extracted a complete set of material
constants $n$, $K$, $(G_{11}-G_{12})$, and $G_{44}$ for the
quantitative description of the MOKE of bcc-Fe(001) at the
laser frequency employed.  To our knowledge this is the first report of the
second order MO-coupling constants of Fe.  They
are comparable to the first-order constants and thus, of general
significance for the theoretical description of the MOKE. The index of
refraction $n$ is in excellent agreement and the linear MO-coupling
constants $K$ agree reasonably with previous works.  In contrast, a
comparison of the second-order constants with earlier QMOKE data from
Postava {\em et al.} demonstrates a remarkable difference.  This
suggests that the anisotropic second-order MO-coupling might strongly
depend on the sample properties.

\section*{Acknowledgements}
The authors would like to thank K. Postava for helpful discussions.

%
%


\begin{thebibliography}{99}
\bibitem{note}
Refs.\ \onlinecite{JAP_91_7293} and \onlinecite{JPD_40_1563} give the
size of QMOKE originating from $m_l m_t$ in saturation, which is a
factor of 2 smaller than our definition of the $q_1$ parameter.
\end{thebibliography}
\end{document}